\documentclass[12pt]{article}
\usepackage{graphicx}

\def\pbnr{}

\def\title{Meson loop singularity, exotic states' productions and decays near threshold}
\def\affiliation{$^a$ Institut f\"{u}r Kernphysik, Institute for Advanced
Simulation, and J\"ulich Center for Hadron
Physics,
    D-52425 J\"{u}lich, Germany\\
       $^b$ Institute of High Energy Physics and Theoretical Physics Center for Science Facilities,
        Chinese Academy of Sciences, Beijing 100049, China}
        \def\support{}
\def\support{This work is supported by DFG and NSFC funds to the Sino-German CRC 110}

\newcommand\pubnumber{\pbnr}
\newcommand\pubdate{\today}
\def\Title#1{\begin{center} {\Large #1 } \end{center}}
\def\Author#1{\begin{center}{ \sc #1} \end{center}}

\newcommand{\OnBehalf}[1]{\sbox0{#1}\ifdim\wd0=0pt
        {}
	\else
	{\\on behalf of #1}
	\fi}
\newcommand{\SupportedBy}[1]{\sbox0{#1}\ifdim\wd0=0pt
        {}
	\else
	{\footnote{#1}}
	\fi}
\def\Address#1{\begin{center}{ \it #1} \end{center}}

\newcommand\pubblock{\includegraphics[width=5cm]{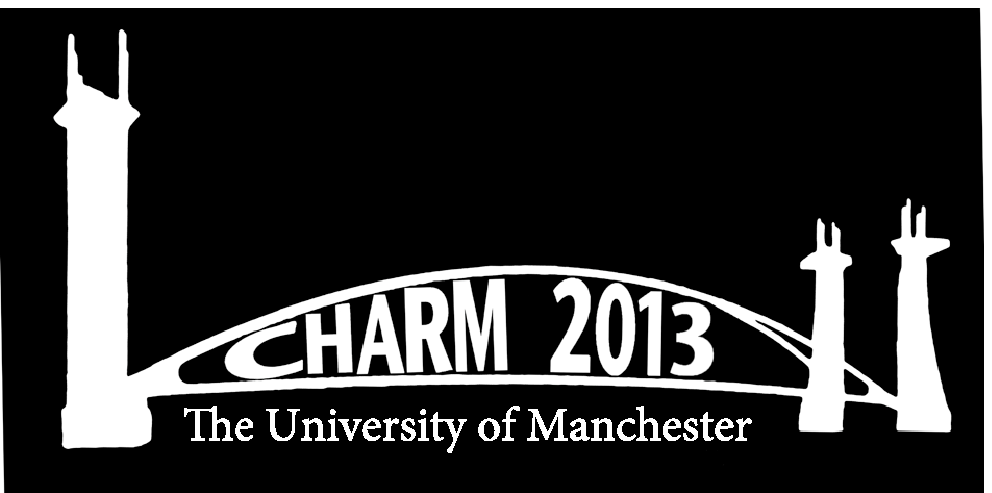}\hfill{\begin{tabular}{l} \pubnumber\\
         \pubdate  \end{tabular}}}
\newenvironment{Abstract}{\begin{quotation}  }{\end{quotation}}
\newenvironment{Presented}{\begin{quotation} \begin{center} 
             PRESENTED AT\end{center}\bigskip 
      \begin{center}\begin{large}}{\end{large}\end{center} \end{quotation}}
\def\Acknowledgements{\bigskip  \bigskip \begin{center} \begin{large}
             \bf ACKNOWLEDGEMENTS \end{large}\end{center}}
\def\venue{The 6$^{th}$ International Workshop on Charm Physics\\
(CHARM 2013)\\
Manchester, UK,  31 August -- 4 September, 2013}

\def\beq{\begin{equation}}
\def\eeq#1{\label{#1}\end{equation}}
\def\eeqn{\end{equation}}

\def\beqa{\begin{eqnarray}}
\def\eeqa#1{\label{#1}\end{eqnarray}}
\def\eeqan{\end{eqnarray}}

\let\bar=\overbar

\def\Dslash{\not{\hbox{\kern-4pt $D$}}}
\def\dslash{\not{\hbox{\kern-2pt $\del$}}}

\def\msb{{\bar{\ssstyle M \kern -1pt S}}}

\begin{document}
\begin{titlepage}
\pubblock

\vfill
\Title{\title}
\vfill
\Author{Qian Wang$^a$\SupportedBy{\support}, Christoph Hanhart$^a$, and Qiang Zhao$^b$}
\Address{\affiliation}
\vfill
\begin{Abstract}
We investigate in detail the ``triangle singularity" regions of meson loops where the corresponding intermediate mesons are nearly-on-shell and have relatively small momenta, and the two heavy mesons strongly interact with each other in an $S$ wave. This $S$-wave interaction will make such kind of meson loops always enhanced and can explain experimental observations of threshold enhancements. We show that the production of 
the $Y(4260)$ and the recently observed $Z_c(3900)$ can be related to each other by this mechanism which will allow for a possible understanding of the nature of these threshold enhancements.

\end{Abstract}
\vfill
\begin{Presented}
\venue
\end{Presented}
\vfill
\end{titlepage}
\def\thefootnote{\fnsymbol{footnote}}
\setcounter{footnote}{0}

\section{Introduction}
During the past decade, a large number of exotic candidates, called $XYZ$ states, have been observed~\cite{Brambilla:2010cs}. Among these states, only a few of them are well established, such as  $X(3872)$,  $X(3915)$, $G(3900)$, $Y(4260)$ and $Y(4360)$. All of those are close to the thresholds of some open flavor channels. For instance, the recently observed $Z_b(10610)$ and $Z_b(10650)$ by the Belle Collaboration~\cite{Belle:2011aa,Adachi:2012im} are close to the $B\bar B^*$ and $B^*\bar B^*$ thresholds, respectively.~\footnote{The notation $B\bar B^*$ means the $B\bar B^*+c.c.$ pair which is also adapted for other analogous cases.} Similarly,  the newly observed $Z_c(3900)$ and $Z_c(4020/4025)$  by BESIII~\cite{Ablikim:2013mio,Ablikim:2013wzq,Ablikim:2013emm}, Belle~\cite{Liu:2013dau} and the analysis of the CLEO-c data~\cite{Xiao:2013iha},  can be viewed as charmonium analogues of $Z_b(10610)$ and $Z_b(10650)$ near the $D\bar D^*$ and $D^*\bar D^*$ threshold, respectively.

For these near threshold states, possible explanations include hadronic molecules or threshold effects which will be the focus of this contribution. For the molecular picture, it contains two conventional mesons, i.e. $Q\bar q$ and $\bar Q q$, as an analogue of the deuteron which is a bound state of proton and neutron. If the pion-exchange interaction between the heavy meson and its anti-heavy meson is strong enough, they can be bound into molecules as first quantitatively discussed by Tornqvist in Ref.~\cite{Tornqvist:1993ng}. For the picture of threshold effects, the kinematic effect becomes important when the incoming and outgoing momentum lie in the ``singularity region"~\cite{Wang:2013hga}. This mechanism does play a role in some kinematic regions no matter the corresponding particles are bound or not.  So it is urgent to identify such singularity kinematics in order to distinguish a genuine state from a threshold cusp effect.

In this contribution, we take the simplest $S$-wave interaction as an example to illustrate this singularity mechanism in Sec.~\ref{sec:singularity}.  As an application, we study the production of the $Z_c(3900)$ at the mass of $Y(4260)$  in Sec.~\ref{sec:Zc3900}. The summary and outlook are given  in the last section.

\section{Analysis of the $S$-wave singularity mechanism}\label{sec:singularity}
\begin{figure}
\begin{minipage}[t]{0.49\linewidth}
\centering
\vspace{-2cm}
\includegraphics[width=2.5in]{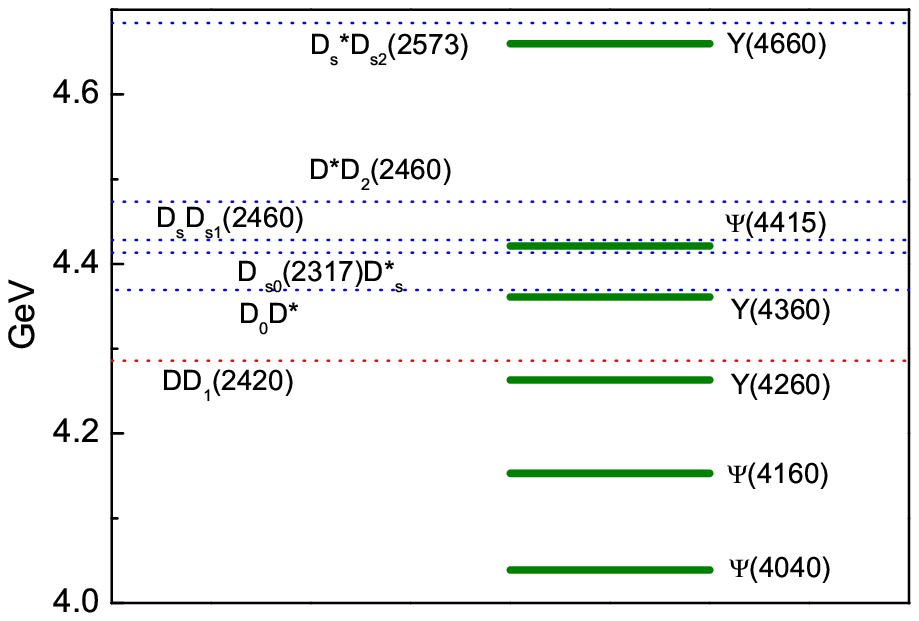}
\vspace{-0.5cm}
\caption{The spectrum of vector charmonium and relative $S$-wave open charm thresholds.}
\label{fig:spectrum}
\end{minipage}\quad
\begin{minipage}[t]{0.49\linewidth}
\centering
\vspace{-2cm}
\includegraphics[width=2.5in]{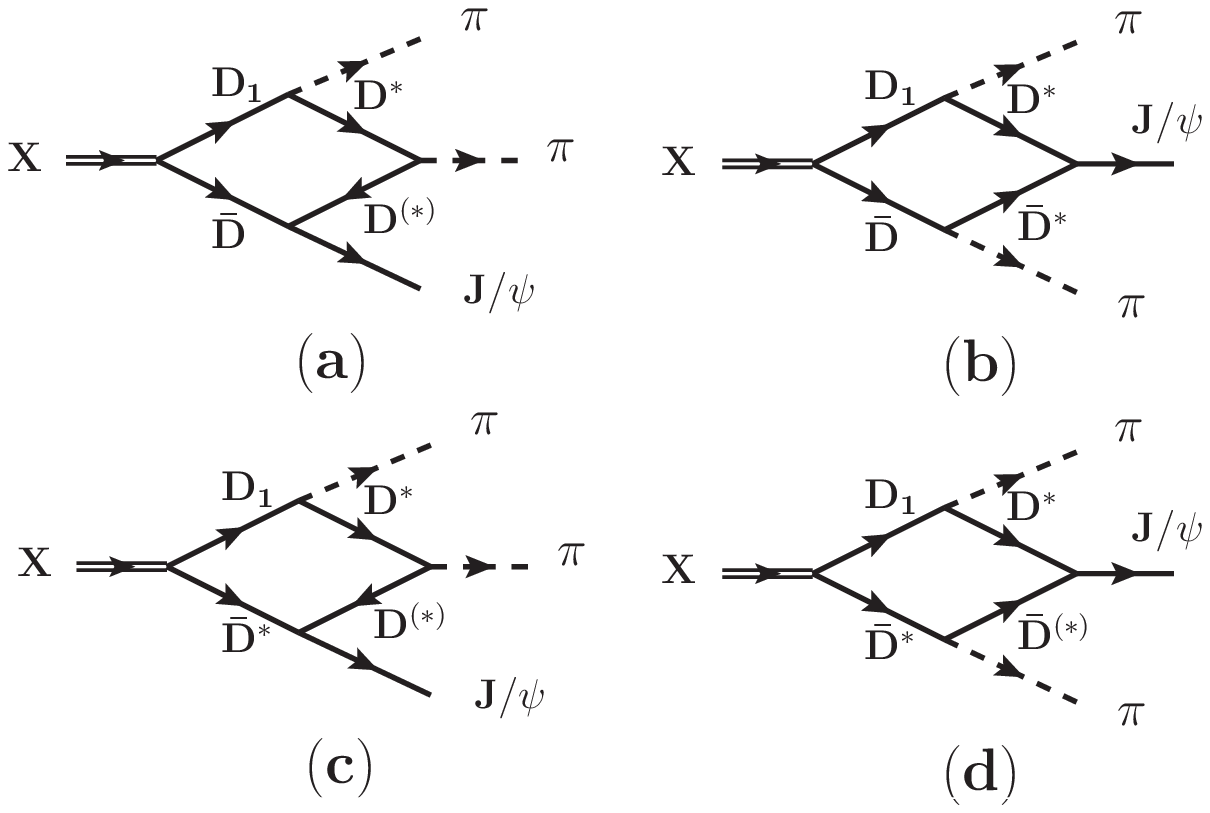}
\caption{Feynman diagrams demonstrating a vector meson $X$ with hidden charm decays into $J/\psi\pi\pi$ via the singularity mechanism.}
\label{fig:Feynman}
\end{minipage}
\end{figure}

In this section, we study the production channel of the $Z_c(3900)$ in the $Y(4260)$ decays to illustrate how this singularity mechanism works. As it is known, the lowest partial wave plays a more important role than other higher partial waves near threshold.  As the first nearby $S$-wave threshold (Fig.~\ref{fig:spectrum}), the $D_1\bar D$ threshold should be crucial in the understanding of the $Y(4260)$ decay. Here, $D_1$ is the narrow state which belongs to the $(\frac{3}{2})^{+}$ spin multiplet in the heavy quark limit. Since $D_1$ mainly decays into $D^*\pi$, a lot of $S$-wave $D\bar D^*$ pairs can be produced and some of them have the probability to form the $Z_c(3900)$~\cite{Wang:2013cya}.

In order to demonstrate the dynamic feature of the relative $S$-wave couplings and low-momentum $D \bar D^*$ scattering, we use the following Lagrangians in our calculation
\begin{eqnarray}\nonumber
  \mathcal{L}_{Y}&=&\frac{y}{\sqrt{2}} Y^{i} \left( D_{1a}^{i\dag} \bar D_a^\dag-D_a^\dag
\bar D_{1a}^{i\dag} \right)+i\frac{y^\prime}{\sqrt 2}\epsilon^{i j k}Y^{\prime i}\left( \bar D_1^{k\dag }D^{*j\dag} - D_1^{k\dag }\bar D^{*j\dag}\right)+H.c.
\label{eq:Lag1}
\end{eqnarray}
for the $Y(4260)$ coupling to the $(\frac{1}{2})^-$ and anti-$(\frac{3}{2})^+$ heavy meson pair, and
\begin{eqnarray}\nonumber
\mathcal{L}_{D_1}&=&i\frac{h^\prime}{f_\pi}
[3D_{1a}^i(\partial^i\partial^j\phi_{ab})D^{*\dag
j}_b-D_{1a}^i(\partial^{j}\partial^j\phi_{ab})D_b^{*\dag
i}\\
&+&3\bar{D}_a^{*\dag
i}(\partial^i\partial^j\phi_{ab})\bar{D}_{1b}^j-\bar{D}_a^{*\dag
i}(\partial^j\partial^j\phi_{ab})\bar{D}_{1b}^i]+H.c.
\end{eqnarray}
for the $D_1$ coupling to the $D^*$ and the pion. The coupling constants $y$ and $h^\prime$ can be found in Refs.~\cite{Casalbuoni:1996pg,Colangelo:2005gb} and the details of other interactions can be found in Ref.~\cite{Cleven:2013sq}.

As shown in Fig.~\ref{fig:Feynman}, no matter what the nature of $X$ is, once it couples to $D_1\bar D^{(*)}$ in the $S$ wave and produces large numbers of $D^*\bar D^{(*)}$ through the $D_1\to D^*\pi$, these diagrams could have significant threshold enhancement when $D_1\bar D^{(*)}$ and  $D^*\bar D^{(*)}$ are allowed to go on-shell simultaneously. Since the exchanged charmed meson between the $J/\psi$ and the $\pi$ is far off-shell, the $D^*\bar D^{(*)}\to J/\psi\pi$ scattering amplitude can be treated as a local function $\mathcal F (M(J/\psi\pi), t)$, with $M(J/\psi\pi)$ and $t$ the invariant mass of $J/\psi\pi$ and $t$-channel momentum transfer, respectively. Since $\mathcal F (M(J/\psi\pi), t)$ does not change very quickly with respect to $M(J/\psi\pi)$ and $t$, the four-point loop function in Fig.~\ref{fig:Feynman} can be written as
\begin{eqnarray}\nonumber
M&=&\int \frac{d^4 l}{(2\pi)^4}\frac{G\epsilon_{X}^i\epsilon_{J/\psi}^j(3q_1^iq_1^j-|q_1|^2\delta^{ij})\mathcal{F}(M(J/\psi\pi),t)}
{(l^0-\frac{|\vec{l}|^2}{2m_{D_1}}+i\varepsilon)(p^0-l^0-\frac{|\vec{l}|^2}{2m_{D^{(*)}}}+i\varepsilon)(l^0-q_1^0-\frac{|\vec{l}-\vec{q_1}|}{2m_{D^*}}+i\varepsilon)}\nonumber\\\nonumber
&\equiv &G\epsilon_{X}^i\epsilon_{J/\psi}^j(3q_1^iq_1^j-|q_1|^2\delta^{ij})\mathcal{F}(M(J/\psi\pi),t) \mathrm{I}(m_{D_1},m_{D^{(*)}},m_{D^*},W,M(J/\psi\pi),m_\pi) \ ,
\end{eqnarray}
where $q_1$ is the three momentum of the final pion connected to the $D_1$, and $I$ is the three-point scalar function.

\begin{figure}
\begin{minipage}[t]{0.49\linewidth}
\centering
\vspace{-0.5cm}
\includegraphics[width=3in]{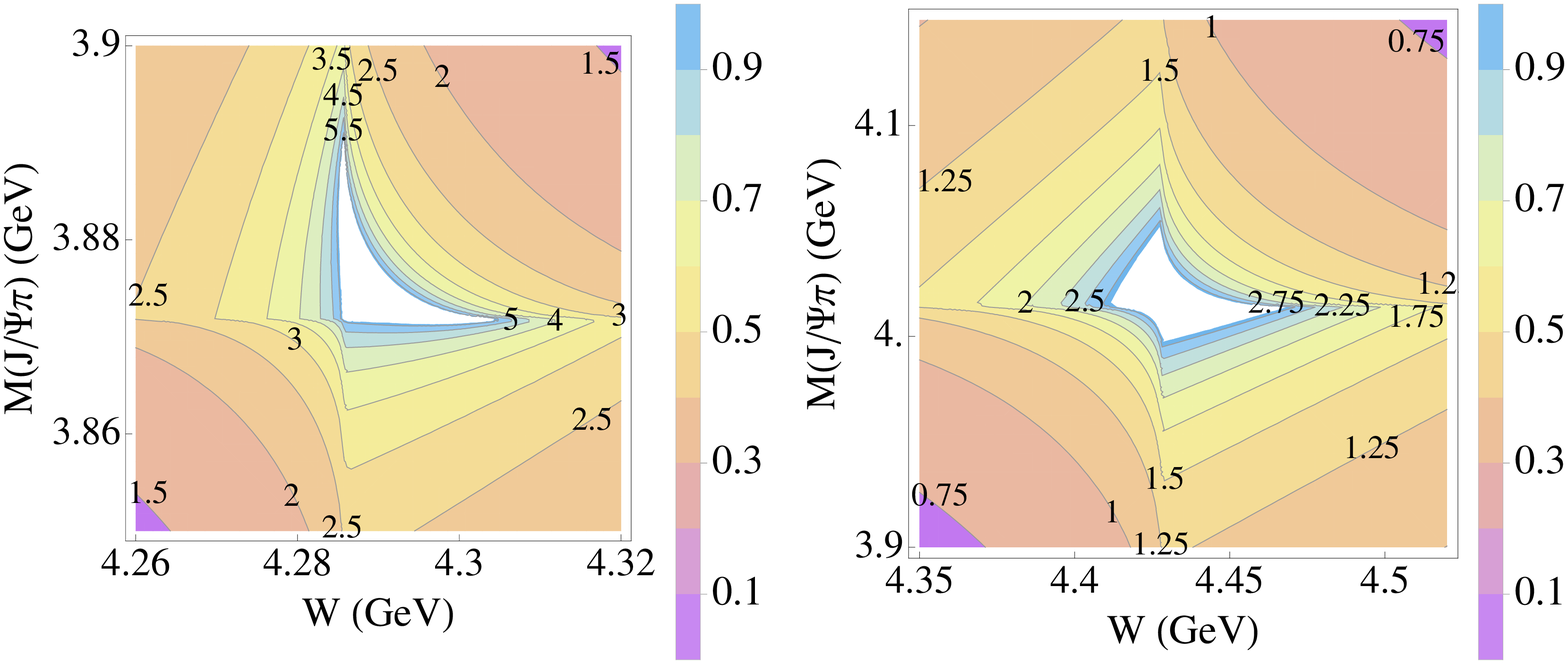}
\caption{The singularity region of $D \bar D^*$ (left panel) and $ D^* \bar D^*$ (right panel) without considering the widths of the intermediated mesons.The numbers in the figures are the absolute values of three-point scalar functions and the numbers in the sidebar are their relative strengths.}
\label{fig:singularity}
\end{minipage}\quad
\begin{minipage}[t]{0.49\linewidth}
\centering
\vspace{-0.5cm}
\includegraphics[width=3in]{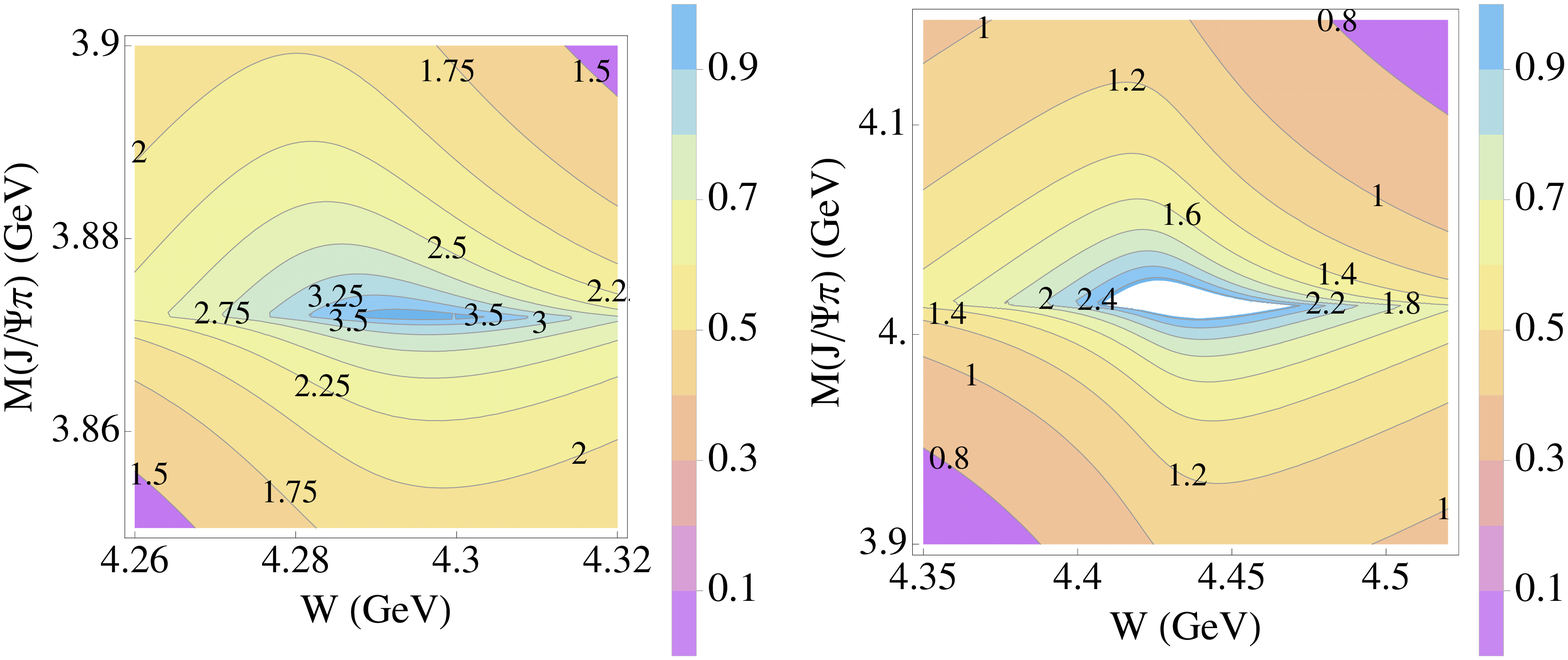}
\caption{The singularity region in the charm sector after considering the widths of the intermediate $D_1$ and $D^*$. The left panel and the right panel are for $D \bar D^*$ and $D^* \bar D^*$ singularity region respectively. The numbers have the same meanings as those in Fig.3.}
\label{fig:singularityWithWidth}
\end{minipage}
\end{figure}

In Fig.~\ref{fig:singularity}, we show contour plots to illustrate the singularity region in the $W-M(J/\psi\pi)$ plane. If the transition amplitude is strongly enhanced by the singularity mechanism, a pronounced cusp will be identified at (or very close to) the $D^*\bar D^{(*)}$ threshold.  From Fig.~\ref{fig:singularity}, we see that a pronounced cusp appears around the $D\bar D^*$ threshold in the c.m. energy region $4.28~\mathrm{GeV}< W < 4.31~\mathrm{GeV}$ (left panel) and another one around the $D^*\bar D^*$ threshold in the region $4.40~\mathrm{GeV}<W<4.45~\mathrm{GeV}$ (right panel). That is because the first cut ($D_1\bar D^{(*)}$) and the second cut ($D^*\bar D^{(*)}$) are satisfied simultaneously in these two kinematic regions which then lead to the triangle singularity being fully operative.  Interestingly, since the mass difference between $D$ and $D^*$ is too large, there is no overlap between these two singularity regions in terms of the c.m. energy. It means we would not expect to see the $D\bar D^*$ and $D^*\bar D^*$ cusps at the same c.m. energy.  We also show the singularity regions after considering the widths of $D_1$ and $D$~\cite{Wang:2013hga} in Fig.~\ref{fig:singularityWithWidth}. The cusp effects at both $D\bar D^*$ and $D^*\bar D^*$ threshold appear smeared significantly. That is also the reason why we only consider the narrow $D_1$ here.

\begin{figure}
\begin{minipage}[t]{0.49\linewidth}
\centering
\vspace{-1cm}
\includegraphics[width=2in]{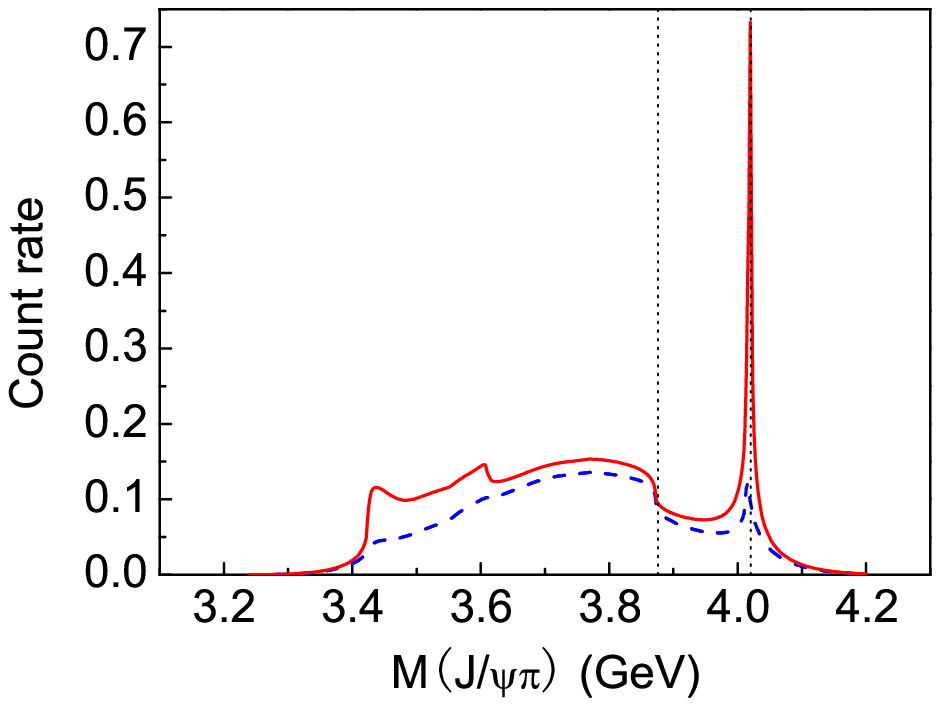}
\vspace{-0.5cm}
\caption{The $J/\psi\pi$ invariant mass distribution at the c.m.
energy $4.43~\mathrm{GeV}$ in the $J/\psi \pi\pi$ channel {\it with}
(dashed) and {\it without} (solid) the width effects for the
intermediate particles. The two vertical dotted lines denote the
$\bar DD^*$ and $\bar D^*D^*$ thresholds, respectively, from left to
right.}
\label{fig:Jpsipi}
\end{minipage}\quad
\begin{minipage}[t]{0.49\linewidth}
\centering
\vspace{-1.5cm}
\includegraphics[width=3in]{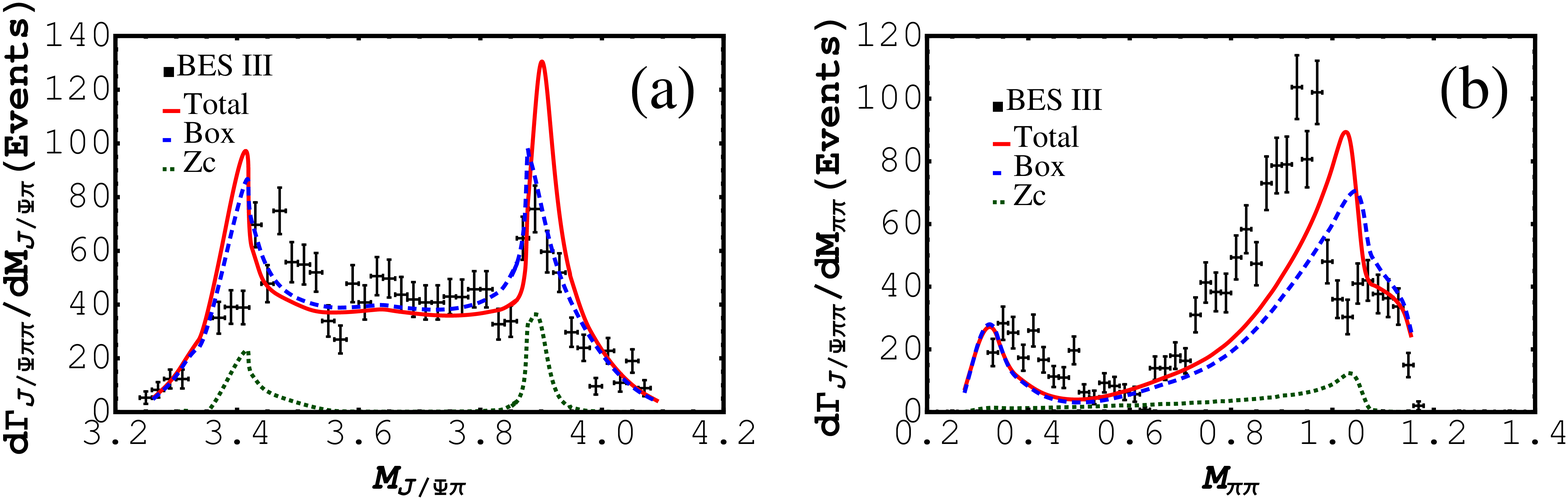}
\vspace{-4cm}
\caption{The invariant mass spectra for (a) $J/\psi\pi$ and (b)
$\pi\pi$ in $Y(4260)\to J/\psi\pi\pi$. The solid, dashed, and
dotted lines stand for the results of the full
calculation, box diagrams, and triangle diagram~\cite{Wang:2013cya} with the $Z_c(3900)$ pole,
respectively.}
\label{fig:YZc3900}
\end{minipage}
\end{figure}

To illustrate the discussions above, we show the $J/\psi\pi$ invariant mass at the c.m. energy $W=4.43~\mathrm{GeV}$ in Fig.~\ref{fig:Jpsipi}.  Since the c.m. energy $4.43~\mathrm{GeV}$ allows the on-shell condition for the $D^*\bar D^*$  instead of the $D\bar D^*$, only one significant cusp effect at the $D^*\bar D^*$ threshold shows up. Comparing to the red solid line which does not consider the widths of the intermediate mesons, the cusp effect of the blue dashed one  (after considering the widths of the intermediate mesons) is  smooth and insignificant.
\section{Production of the $Z_c(3900)$ in the $Y(4260)$ decay}\label{sec:Zc3900}
In this Section we present quantitative results for the production of the $Z_c(3900)$ in the $Y(4260)$ decay where the triangle singularity will play a crucial role~\cite{Wang:2013cya}. 
Because the c.m. energy $4.26~\mathrm{GeV}$ lies in the $D\bar D^*$ singularity region, copious $S$-wave $D\bar D^*$ will be produced and a clear cusp effect in the $J/\psi\pi$ invariant mass distribution will be expected. As shown in Fig. 1 in Ref.~\cite{Wang:2013cya}, besides the box diagram contributions, we also include contributions of the $Z_c(3900)$. After considering the $\pi\pi$ final state interaction (FSI) properly, the contributions of the box diagram, the $Z_c(3900)$ pole and the sum of them are shown in Fig.~\ref{fig:YZc3900} individually.   Since the kinematic region of the two pion invariant mass is from the two-pion threshold to more than $1~\mathrm{GeV}$ and they are in the isoscalar state, the $S$-wave $\pi\pi$ interaction plays a more important role. As a result, we only consider the $S$-wave $\pi\pi$ interaction here.

The numerical results for the $J/\psi\pi$ and $\pi\pi$
invariant mass spectra of $Y(4260)\to J/\psi\pi\pi$ are shown in
Fig.~\ref{fig:YZc3900}.  The dashed, dotted and solid lines denote the results from the box diagrams, $Z_c(3900)$ pole diagrams and the sum of them, respectively. In our case the box diagrams play a dominate role compared to the $Z_c(3900)$ pole diagrams. However, as shown by the dashed lines, an explicit enhancement around 3.9 GeV in the $J/\psi\pi$ spectrum can be produced because of the nearly-on-shell two-cut singularity condition as discussed in the previous Section. The bump structure at about $3.4~\mathrm{GeV}$ is the kinematic reflection of the $D\bar D^*$ threshold enhancement. The explicit inclusion of the $Z_c(3900)$ makes the enhancement near the $D\bar D^*$ threshold broader. If the bump structure at $D\bar D^*$ threshold is from the singularity mechanism, it will be sensitive to the incoming energy. So further scans at different c.m. energies especially out of the singularity region are necessary to determine whether the $Z_c(3900)$ is a genuine state or not.

In our scenario, a detailed analysis of the relative partial waves between the two pions demonstrates that in addition to the $S$ wave, the $D$ wave also contributes to the $\pi\pi$ productions. However, the $S$-wave dominance will result in a broad bump at lower invariant mass region and a flattened dip at about $0.5~\mathrm{GeV}$. This behavior is mainly driven by the box diagram. The dip structure in Fig.~\ref{fig:YZc3900} at around $1~\mathrm{GeV}$ should be located exactly at the $K\bar K$ threshold if the $S$-wave partial wave is the only contribution. However, the data show that the dip position is slightly shifted to be higher than 1 GeV. This indicates the presence of other higher partial waves.

Since the invariant mass distributions of the $J/\psi\pi$ and $\pi\pi$ can be well explained if the $Y(4260)$ couples strongly to $D_1\bar D$, we can expect that the $Y(4260)$ should be dominated by the $D_1\bar D$ component. If this is the case, we expect that $D\bar D^*\pi$ should be the dominate decay mode through the intermediate $D_1\to D^*\pi$ which will explain the large deficit between the total width and its decay into $J/\psi\pi\pi$.  We also expect an asymmetric spectral shape of the $Y(4260)$ due to the nearby $D_1\bar D$ threshold~\cite{Cleven:2013mka}. The $D_1\bar D$ molecule scenario also predicts nontrivial cross section line shape for the $J/\psi\pi\pi$ and $h_c\pi\pi$ productions around the $Y(4260)$ for which a detailed study can be found in Ref.~\cite{Cleven:2013mka}.
\section{Summary and Outlook}\label{sec:summary}
We identify the triangle singularity kinematic regions in the heavy quarkonium decays involving $S$-wave vertices.
As a result, the $(\frac{1}{2})^-$ and $(\frac{3}{2})^+$ open charm threshold play an important role for certain vector charmonium decays.
There are some kinematic regions that can fulfill the two-cut condition such
that the intermediate heavy meson loop can produce significant cusp effects and enhance the transition amplitude significantly.
The clarification of the origin of the cusps and their evolutions with the initial masses
would be important for our understanding of those near threshold states, such as $X(3872)$, $Y(4260)$, $Z_c(3900)$ and so on.
We emphasize that a genuine state can also be observed even out of the singularity regions
defined in this work. This will provide more information about the nature of those threshold states.
Without introducing any strong assumption,  the molecular nature of $Y(4260)$ as a bound state of $D_1\bar D$  can naturally explain the
observation of an enhancement around $D\bar D^*$ threshold in the $J/\psi\pi$
invariant mass spectrum. This scenario describes the experimental data very well
and provides a strong evidence for the molecular nature of $Y(4260)$.

\Acknowledgements
Q. Wang acknowledges the collaborations with Martin Cleven, Feng-Kun Guo, Ulf-G. Mei\ss ner and Xiao-Gang Wu on the works presented in this conference and many relevant discussions and works which cannot be covered in this proceeding.

\end{document}